\documentclass[prl,twocolumn,aps]{revtex4}

\usepackage{graphicx} 

\newcommand{\ket}[1]{\left|#1\right\rangle}

\begin{document}

\title{Proof of finite surface code threshold for matching}

\author{Austin G. Fowler}
\affiliation{Centre for Quantum Computation and Communication
Technology, School of Physics, The University of Melbourne, Victoria
3010, Australia}

\date{\today}

\begin{abstract}
The field of quantum computation currently lacks a formal proof of experimental feasibility. Qubits are fragile and sophisticated quantum error correction is required to achieve reliable quantum computation. The surface code is a promising quantum error correction code, requiring only a physically reasonable 2-D lattice of qubits with nearest neighbor interactions. However, existing proofs that reliable quantum computation is possible using this code assume the ability to measure four-body operators and, despite making this difficult to realize assumption, require that the error rate of these operator measurements is less than $10^{-9}$, an unphysically low target. High error rates have been proved tolerable only when assuming tunable interactions of strength and error rate independent of distance, which is also unphysical. In this work, given a 2-D lattice of qubits with only nearest neighbor two-qubit gates, and single-qubit measurement, initialization, and unitary gates, all of which have error rate $p$, we prove that arbitrarily reliable quantum computation is possible provided $p<7.4\times 10^{-4}$, a target that many experiments have already achieved. This closes a long-standing open problem, formally proving the experimental feasibility of quantum computation under physically reasonable assumptions.
\end{abstract}

\maketitle

\begin{figure}
\begin{center}
\resizebox{60mm}{!}{\includegraphics{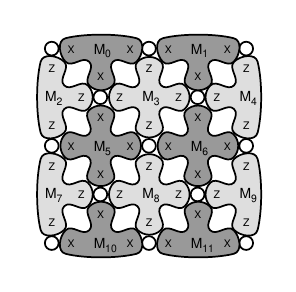}}
\end{center}
\caption{A small surface code. Larger codes can be constructed by expanding the pattern. Circles represent qubits. Each $M_i$ represents an operator (tensor product of Pauli $X$ or $Z$ operators) that is measured to detect errors. Note that all operators commute.}\label{sc}
\end{figure}

A 2-D array of qubits with tunable nearest neighbor interactions is a believable experimental target \cite{Devi08,Amin10,Jone10,Kump11}. The surface code \cite{Brav98,Denn02,Raus07,Raus07d,Fowl08,Fowl12f} can be implemented optimally using such an array. Qubits undergoing very general unwonted evolution can be accurately modeled as suffering random Pauli $X$ and $Z$ errors \cite{Shor95}. This is a reasonable model as quantum error correction (QEC) involves repeated operator measurements that project noisy qubits onto states that differ from the desired state by Pauli operators. An expandable pattern of operator measurements $M_i$ associated with the surface code is shown in Fig.~\ref{sc}. The operators $M_i$ are called stabilizers \cite{Gott97}.

Let the state of the qubits in Fig.~\ref{sc} be $\ket{\Psi}$ and, without loss of generality, let us assume that we have measured all stabilizers $M_i$ and observed the +1 eigenstate, so $M_i\ket{\Psi}=\ket{\Psi}$ for all $i$. Note that $[M_i,M_j]=0$ for all $i$, $j$, so such a state exists. Suppose the central qubit suffers some general error $E$, where
\begin{eqnarray}
E & = & \left(\begin{array}{cc} a & b \\ c & d \end{array}\right) \\
  & = & \frac{a+d}{2}I + \frac{b+c}{2}X + \frac{-b+c}{2}XZ + \frac{a-d}{2}Z. \label{decomp}
\end{eqnarray}
Eq.~\ref{decomp} shows that $E$ is a linear superposition of no error ($I$), or an $X$, $Z$ or $XZ$ error. After the error $E$, a subsequent round of $M_i$ measurements will project $E$ to $I$ if $M_i=+1$ for $i=3,5,6,8$, $X$ if $M_i=+1$ for $i=5,6$ and $M_i=-1$ for $i=3,8$, $Y$ if $M_i=-1$ for $i=3,5,6,8$, and $Z$ if $M_i=+1$ for $i=3,8$ and $M_i=-1$ for $i=5,6$. All measurements $M_i$, $i\neq 3,5,6,8$, will remain +1.

The earliest proof that arbitrary reliability could be achieved using the surface code assumed that each measurement $M_i$ had a probability $q$ of reporting the wrong result and that between each round of measurement each qubit suffered a $Z$ error with probability $p$ \cite{Harr04}. The proof showed that arbitrary reliability was achievable given sufficient qubits for $p+q<2.4\times 10^{-11}$. More recently, assuming perfect stabilizer measurements and a 3-D array of qubits each suffering a single error with probability $p$, assumptions that are equivalent to a 2-D array of qubits with periodic faulty measurements and qubit errors, arbitrary reliability was proved possible given sufficient qubits for $p<1.4\times 10^{-9}$ \cite{Brav11}. These error bounds are unphysically low.

Error rates slightly above $10^{-3}$ have been proved tolerable in other quantum error correction codes when assuming tunable interactions between arbitrary pairs of qubits with interaction time and error rate independent of the physical separation of the qubits \cite{Alif07c,Paetznick2011}. However, no physical machine possesses such interactions. As such, there is currently no formal proof that arbitrarily reliable quantum computation is experimentally feasible, raising serious questions about the validity of quantum computation. In this work, we provide the much-needed formal proof of experimental feasibility.

It has long been believed that error rates of $10^{-4}$ are a reasonable experimental target \cite{Pres98b,Gaeb12}. Experimentally, single-qubit measurement with error rate $10^{-4}$ and initialization with error $10^{-5}$ have been demonstrated \cite{Myer08}. Single-qubit unitary gates have been demonstrated with error $2\times 10^{-5}$ \cite{Brow11}. Two-qubit gates are the most technically challenging, with Bell state preparation with error $7\times 10^{-3}$ the current state-of-the-art \cite{Benh08}. There is no physical reason to believe the technical challenges cannot be overcome and similarly low-error two-qubit gates achieved. All of these experiments were performed using ion traps \cite{Kiel02}, a technology well-suited to implementing a 2-D array of qubits with tunable nearest neighbor interactions \cite{Amin10}. We therefore seek a formal proof that arbitrarily reliable quantum computation can be achieved given a 2-D array of qubits with only nearest neighbor two-qubit gates, single-qubit measurement, initialization, and unitary gates, with all gates having an error rate $p\sim 10^{-4}$.

\begin{figure}
\begin{center}
\resizebox{60mm}{!}{\includegraphics{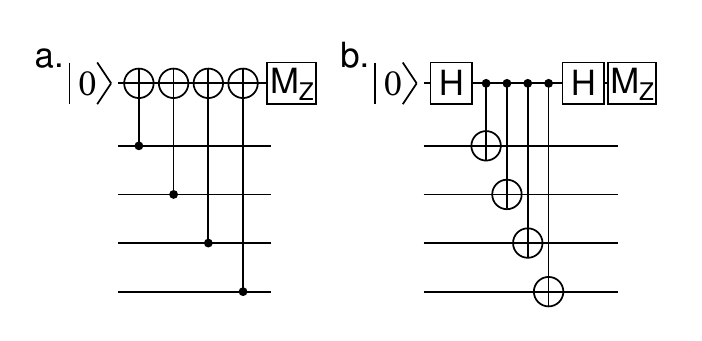}}
\end{center}
\caption{Quantum circuits measuring {\bf a.}~ZZZZ and {\bf b.}~XXXX operators. The $\ket{0}$ represents initialization, the $H$ represents Hadamard, the $M_Z$ represents measurement of the operator $Z$, and the dot and target symbols connected by lines each represent a $C_X$ gate. The $C_X$ inverts the value of the target qubit if the control (dot) qubit is in the state $\ket{1}$ and does nothing otherwise. For example, $C_X(\alpha\ket{0}+\beta\ket{1})\ket{0}=\alpha\ket{00}+\beta\ket{11}$. The interaction sequence is North, West, East, South, as shown in Fig.~\ref{syn_err}.}\label{syn_circs}
\end{figure}

The only nontrivial unitary gates we will use will be Hadamard $H$ ($H\ket{0}=\ket{+}$, $H\ket{1}=\ket{-}$) and controlled-NOT $C_X$ ($C_X\ket{10}=\ket{11}$, $C_X\ket{11}=\ket{10}$). Quantum circuits measuring the stabilizers $M_i$ are shown in Fig.~\ref{syn_circs}. Note that these circuits can be implemented in parallel across an arbitrarily large surface. A single round of error detection is defined to be a single parallel execution of these circuits. Initialization to $\ket{0}$ results in $\ket{1}$ with probability $p$, measurement reports the wrong eigenstate with probability $p$, $H$ and identity $I$ are followed by an error randomly chosen from $X$, $Y$, $Z$ with probability $p$, and $C_X$ is followed by an error randomly chosen from $IX$, $IY$, $IZ$, $XI$, ..., $ZZ$ with probability $p$. This error model is justified by the above discussion of arbitrary errors being projected to Pauli errors.

Errors followed by even perfect $C_X$ gates can result in multiple errors. For example, $C_X\ket{00}=\ket{00}$, however if there is an $X_1$ error then $C_XX_1\ket{00}=C_X\ket{10}=\ket{11}=X_1X_2\ket{00}$. In words, the $C_X$ copies $X$ errors on the control qubit to the target qubit. Similarly, $C_X\ket{++}=\ket{++}$,  however if there is a $Z_2$ error then $C_XZ_2\ket{++}=C_X\ket{+-}=\ket{--}=Z_1Z_2\ket{++}$. In words, the $C_X$ gate copies $Z$ errors on the target qubit to the control qubit.

\begin{figure}
\begin{center}
\resizebox{80mm}{!}{\includegraphics{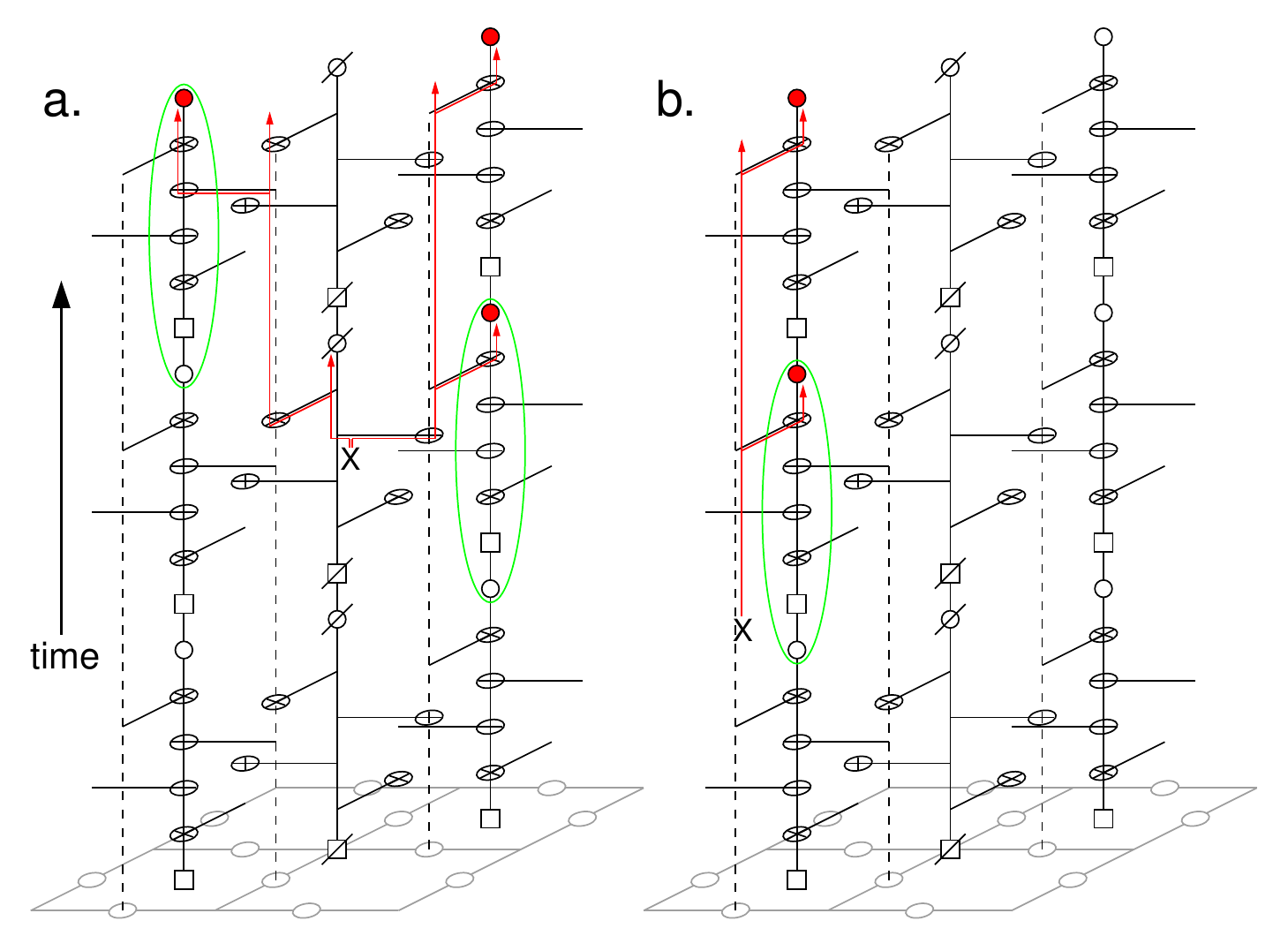}}
\end{center}
\caption{(Color online) 2-D surface code (grey). Time runs vertically. Squares represent initialization to $\ket{0}$, circles represent $Z$ basis measurement. Slashed squares represent initialization to $\ket{+}$, slashed circles represent $X$ basis measurement (both achieved using Hadamard gates). {\bf a.}~A single error leading to a pair of detection events (long vertical ellipses). A detection event is a sequential pair of measurements with differing value. Lines with arrows show the branching paths of error propagation. {\bf b.}~An error leading to a single detection event due to proximity to a boundary of the lattice.}\label{syn_err}
\end{figure}

Define a detection event to be a pair of sequential stabilizer measurements that differ in value. Fig.~\ref{syn_err}a shows a single $X$ error propagating through rounds of surface code error detection circuitry until two detection events are generated. $X$ errors are detected by sequential pairs of $Z$ stabilizer measurements, $Z$ errors by sequential pairs of $X$ stabilizer measurements. Given the right pattern of errors, any pair of sequential stabilizer measurements can be associated with a detection event. We associate a specific space-time location with each potential detection event, namely the space-time midpoint between sequential measurements.

\begin{figure}
\begin{center}
\resizebox{80mm}{!}{\includegraphics{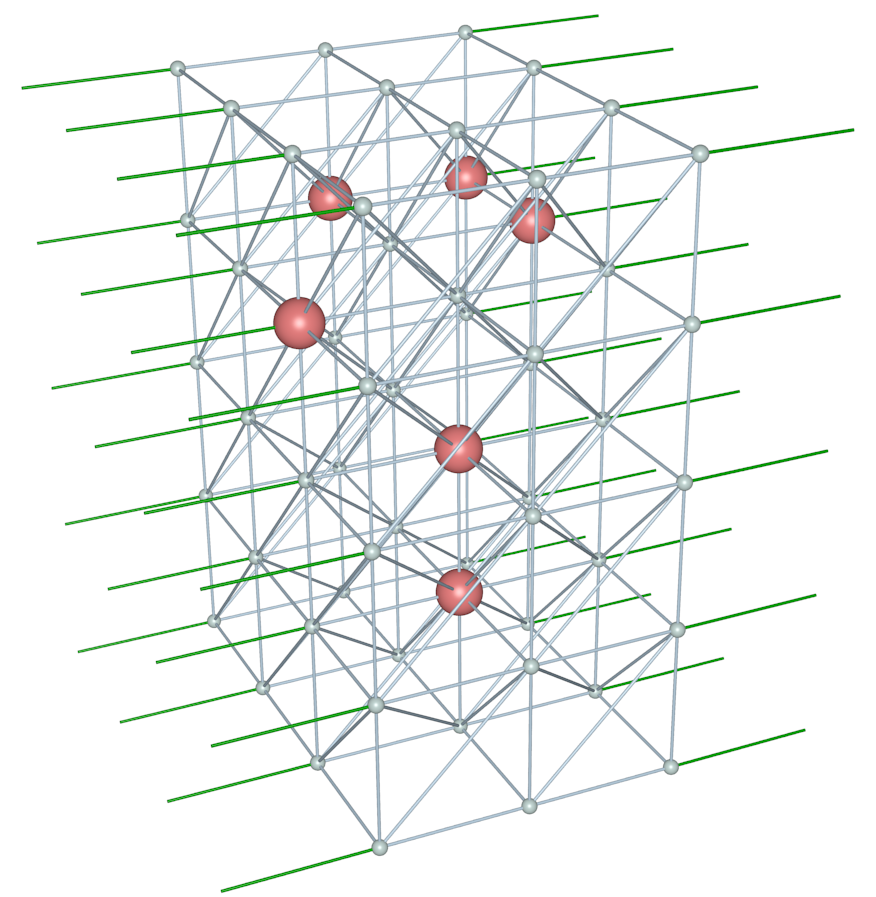}}
\end{center}
\caption{(Color online) Distance 4 example of a lattice of dots and lines with stochastically generated detection events. Dots (small spheres) correspond to space-time locations where the endpoints of error chains could potentially be detected. Detection events (large spheres) correspond to space-time locations where error chain end points have been detected. Lines link pairs of dots where a pair of detection events could potentially be generated by a single error. Darker lines link spatial boundaries to a single dot where a single detection event could be generated by a single error.}\label{problem}
\end{figure}

For visualization purposes, we draw a sphere at every space-time location a detection event can occur. We draw a cylinder between every pair of space-time locations that can be associated with detection events generated by a single error. We call this structure of spheres and cylinders a lattice of dots and lines. Lattices are constructed by studying the propagation of all errors through the periodic surface code quantum circuit \cite{Wang11,Fowl12d}. An example of a surface code lattice is shown in Fig.~\ref{problem}, including detection events stochastically generated by errors during a simulation of the surface code. Note the lines leading to nowhere, indicating the nearby presence of potential errors leading to single detection events. This only occurs near boundaries of the array of qubits. Note also the regular yet nontrivial structure of the lines.

There are two lattices, one associated with detecting $X$ errors and one with detecting $Z$ errors. The distance of a surface code is defined to the length in lines of the shortest path between disjoint boundaries of a lattice. Paths of lines connecting disjoint boundaries can be associated with chains of Pauli operators that commute with all stabilizers yet are not products of stabilizers themselves. Such chains of operators are called logical operators and manipulate the data stored in the surface code.

For the purposes of this proof, we associate a weight of 1 with each line. Given a lattice with random detection events, a minimum weight perfect matching is a set of paths through the lattice such that every detection event is incident on exactly one path, paths are allowed to terminate at boundaries, and the total weight of all paths is minimal. Edmonds' minimum weight perfect matching algorithm \cite{Edmo65a,Edmo65b,Fowl11b,Fowl12c} can efficiently find such a set of paths. Other algorithms have been used to correct errors in the surface code \cite{Ducl09,Ducl10,Bomb12,Woot12}, however matching currently remains the only algorithm that has been used to efficiently handle errors in implementations of the surface code making use of only two-qubit interactions and single qubit measurements.

Define the line probability $\epsilon$ to be the total probability of all single errors leading to detection events at the endpoints of the line. Given Pauli error models for each quantum gate, with the strength of each error model scaled by a global parameter $p$, the probability of a given line can be expressed as a polynomial in $p$, with a linear relationship at low $p$ \cite{Wang11}.

Let $V$ be an $n\times n\times n$ dot volume of lattice with non-cyclic boundaries. Assume each pair of opposing faces is a boundary of a distinct type. Let $p$ be a characteristic physical gate error rate low enough to ensure the probability of any given line in $V$ being associated with an error is less than some small $\epsilon$. For minimum weight perfect matching to fail to correct the errors in $V$, there must be at least one path of $m\geq n+1$ lines connecting opposing boundaries containing at least $\lceil m/2 \rceil$ lines associated with errors. A path corresponding to a logical error not containing at least $\lceil m/2 \rceil$ lines associated with errors can be matched with lower weight, contradicting the assumption of a minimum weight perfect matching.

The number of paths of length $m$ can be upper bounded by choosing three of the six faces of $V$ as path starting points. This results in $3n^2$ starting points. Each dot is connected to at most 12 neighboring dots (Fig.~\ref{problem}). A general path from dot to dot that does not backtrack on itself must link to one of 11 of the surrounding 12 dots (the 12th is already part of the path). After the starting point is chosen, there are only $m-1$ direction decisions to make. There are thus no more than $3n^2 11^{m-1}$ paths of length $m$. Since we are only interested in non-self-intersecting, opposing boundary connecting paths, this is a generous upper bound.

Given a particular path of length $m$, the probability of at least $\lceil m/2 \rceil$ of its lines being associated with errors is
\begin{eqnarray}
\sum_{i=\lceil \frac{m}{2} \rceil}^m \left( \begin{array}{c} m \\ i \end{array} \right)\epsilon^i & \leq & \sum_{i=\lceil \frac{m}{2} \rceil}^m \left( \begin{array}{c} m \\ \lceil \frac{m}{2} \rceil \end{array} \right)\epsilon^i \\
& = & \left( \begin{array}{c} m \\ \lceil \frac{m}{2} \rceil \end{array} \right)\epsilon^{\lceil \frac{m}{2} \rceil}\sum_{i=0}^{m-\lceil \frac{m}{2} \rceil}\epsilon^i \\
& \leq & \left( \begin{array}{c} m \\ \lceil \frac{m}{2} \rceil \end{array} \right)\epsilon^{\lceil \frac{m}{2} \rceil}\frac{1}{1-\epsilon} \\
& \leq & \epsilon^{\lceil \frac{m}{2} \rceil}\sum_{i=0}^m \left( \begin{array}{c} m \\ i \end{array} \right) \\
& = & 2^m \epsilon^{\lceil \frac{m}{2} \rceil}
\end{eqnarray}
The probability of a logical error is therefore no more than
\begin{eqnarray}
& & \sum_{m=n+1}^\infty 3n^2 11^{m-1} 2^m \epsilon^{\lceil \frac{m}{2} \rceil} \\
& = & \frac{3n^2}{11}\sum_{m=n+1}^\infty 22^m \epsilon^{\lceil \frac{m}{2} \rceil} \\
& \leq & \frac{3n^2}{11}\sum_{m=n+1}^\infty 22^m \epsilon^{m/2} \\
& = & \frac{3n^2}{11}\left( 22\sqrt{\epsilon}\right)^{n+1}\sum_{m=0}^\infty \left( 22\sqrt{\epsilon}\right)^m \\
& = & \frac{3n^2}{11}\left( 22\sqrt{\epsilon}\right)^{n+1} \frac{1}{1- 22\sqrt{\epsilon}} \label{expsup}
\end{eqnarray}
Provided $\epsilon < 1/484$, the above can be made arbitrarily small by increasing $n$. Since $\epsilon$ can be expressed as a polynomial in $p$ independent of $n$, this proves that using minimum weight perfect matching to correct errors in the surface code results in a finite threshold error rate. Assuming the surface code circuits and error models in \cite{Wang11}, in which the most errorprone line satisfied $\epsilon < 14p/5$, a lower bound to the threshold error rate of $7.4\times 10^{-4}$ is obtained. Given experimental achievements to date, this is sufficiently high to formally prove the experimental feasibility of arbitrarily reliable quantum computation.

Eq.~\ref{expsup} also proves that logical errors are exponentially suppressed with code distance, implying extremely low logical error rates can be achieved with modest qubit overhead. Furthermore, as our proof is fundamentally based on minimum weight perfect matching and this algorithm is highly efficient \cite{Fowl11b,Fowl12c}, we have proved that the classical computing overhead is also modest.

This research was conducted by the Australian Research Council Centre of Excellence for Quantum Computation and Communication Technology (project number CE110001027), with support from the US National Security Agency and the US Army Research Office under contract number W911NF-08-1-0527. Supported by the Intelligence Advanced Research Projects Activity (IARPA) via Department of Interior National Business Center contract number D11PC20166.  The U.S. Government is authorized to reproduce and distribute reprints for Governmental purposes notwithstanding any copyright annotation thereon.  Disclaimer: The views and conclusions contained herein are those of the authors and should not be interpreted as necessarily representing the official policies or endorsements, either expressed or implied, of IARPA, DoI/NBC, or the U.S. Government.

\bibliography{../References}

\end{document}